\documentclass[doublespacing]{elsart}

\usepackage{icarus}

\usepackage{natbib}
\usepackage{aas_macros}
\usepackage{color}

\bibpunct{(}{)}{;}{a}{,}{,}

\usepackage{graphicx}

\usepackage{amssymb}
\usepackage{amsmath, esint}

\usepackage{hyperref}
\usepackage{placeins}

\newcommand{\BibTeX}{ \textrm{B\kern-.05em\textsc{i\kern-.025em b}\kern-.08em
    T\kern-.1667em\lower.7ex\hbox{E}\kern-.125emX} }
\begin{document}

\begin{frontmatter}



\title{A Framework for Relating the Structures and Recovery Statistics in Pressure Time-Series Surveys for Dust Devils}


\author[Brian]{Brian Jackson}, 
\author[Ralph]{Ralph Lorenz}, 
\author[Brian]{Karan Davis}

\address[Brian]{Department of Physics,
				Boise State University,
				1910 University Drive,
				Boise, ID 83725-1570, USA}
\address[Ralph]{Johns Hopkins University Applied Physics Lab,
				11100 Johns Hopkins Road, Laurel,
				Maryland 20723-6099, USA}


%


\end{frontmatter}



\begin{flushleft}
\vspace{1cm}
Number of pages: \pageref{lastpage} \\
Number of figures: \ref{fig:integration_path}\\
\end{flushleft}


\begin{pagetwo}{De-biasing Dust Devil Surveys}

Brian Jackson\\
Department of Physics,\\
Boise State University\\
1910 University Drive\\
Boise, ID 83725-1570, USA\\
\\
Email: bjackson@boisestate.edu\\
Phone: (208) 426-3723 \\
\end{pagetwo}

\begin{abstract}
Dust devils are likely the dominant source of dust for the martian atmosphere, but the amount and frequency of dust-lifting depend on the statistical distribution of dust devil parameters. Dust devils exhibit pressure perturbations and, if they pass near a barometric sensor, they may register as a discernible dip in a pressure time-series. Leveraging this fact, several surveys using barometric sensors on landed spacecraft have revealed dust devil structures and occurrence rates. However powerful they are, though, such surveys suffer from non-trivial biases that skew the inferred dust devil properties. For example, such surveys are most sensitive to dust devils with the widest and deepest pressure profiles, but the recovered profiles will be distorted, broader and shallow than the actual profiles. In addition, such surveys often do not provide wind speed measurements alongside the pressure time series, and so the durations of the dust devil signals in the time series cannot be directly converted to profile widths. Fortunately, simple statistical and geometric considerations can de-bias these surveys, allowing conversion of the duration of dust devil signals into physical widths, given only a distribution of likely translation velocities, and the recovery of the underlying distributions of physical parameters. In this study, we develop a scheme for de-biasing such surveys. Applying our model to an in-situ survey using data from the Phoenix lander suggests a larger dust flux and a dust devil occurrence rate about ten times larger than previously inferred. Comparing our results to dust devil track surveys suggests only about one in five low-pressure cells lifts sufficient dust to leave a visible track.
\end{abstract}


\section{Introduction}
\label{sec:introduction}
Dust devils are small-scale (a few to tens of meters) low-pressure vortices rendered visible by lofted dust. They usually occur in arid climates on the Earth and ubiquitously on Mars. Martian dust devils have been studied with orbiting and landed spacecraft and were first identified on Mars using images from the Viking Orbiter \citep{Thomas_1985}. On Mars, dust devils may dominate the supply of atmospheric dust and influence climate \citep{Basu_2004}, pose a hazard for human exploration \citep{Balme_2006}, and have lengthened the operational lifetime of Martian rovers \citep{Lorenz_Reiss_2014}. On the Earth, dust devils significantly degrade air quality in arid climates \citep{Gillette_1990} and may pose an aviation hazard \citep{Lorenz_2005}.

The dust-lifting capacity of dust devils seems to depend sensitively on their structures, in particular on the pressure wells at their centers \citep{Neakrase_2006}, so the dust supply from dust devils on both planets may be dominated by the seldom observed, most vigorous devils. Thus, elucidating the origin, evolution, and population statistics of dust devils is critical for understanding important terrestrial and Martian atmospheric properties. 

Studies of Martian dust devils have been conducted through direct imaging of the devils and identification of their tracks on Mars' dusty surface \citep{Balme_2006}. Studies with in-situ meteorological instrumentation have also identified dust devils, either via obscuration of the Sun by the dust column \citep{Zorzano_2013} or their pressure signals \citep{Ellehoj_2010}. Studies have also been conducted of terrestrial dust devils and frequently involve in-person monitoring of field sites. Terrestrial dust devils are visually surveyed \citep{Pathare_2010}, directly sampled \citep{Balme_2003}, or recorded using in-situ meteorological equipment \citep{Sinclair_1973,Lorenz_2012}.

As noted in \citet{Lorenz_2009}, in-person visual surveys are likely to be biased toward detection of larger, more easily seen devils. Such surveys would also fail to recover dustless vortices \citep{Lorenz_2015}. Recently, terrestrial surveys similar to Martian dust devil surveys have been conducted using in-situ single barometers \citep{Lorenz_2012,Lorenz_2014,Jackson_2015} and photovoltaic sensors \citep{Lorenz_2015}. These sensor-based terrestrial surveys have the advantage of being directly analogous to Martian surveys and are cost-effective compared to the in-person surveys.

\begin{figure}
\begin{center}
\includegraphics[width=\textwidth]{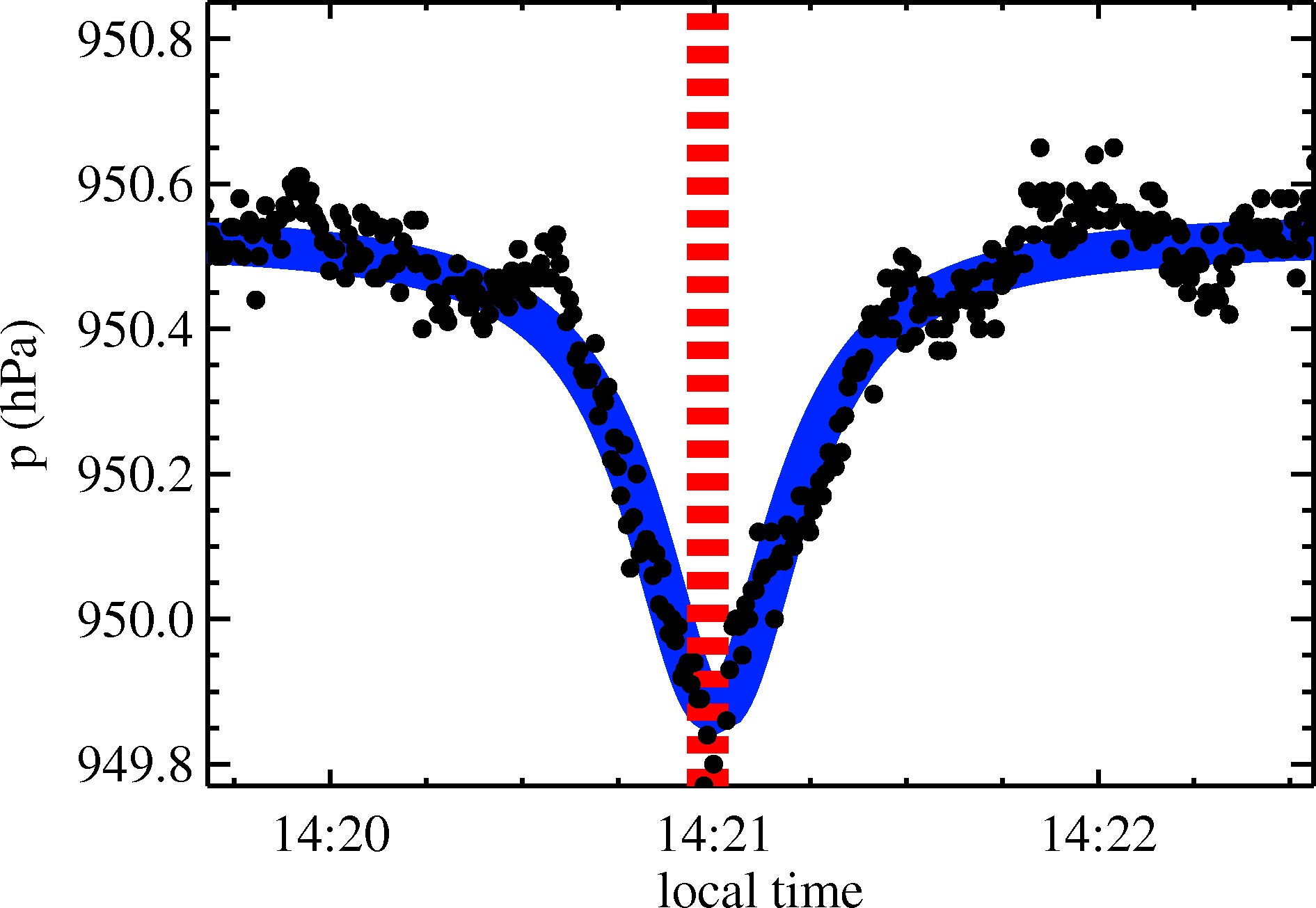}
\caption{Example dust devil profile from \citet{Jackson_2015}. The black dots show the pressure $p$ measurements (in hectoPascal hPa) as a function of local time in hours. The blue line shows the best-fit profile model, and the red, dashed line the profile center. Both lines are made extra thick only to enhance visibility.}
\label{fig:conditioning_detection_b_inset}
\end{center}
\end{figure}

Single-barometer surveys have been successful on both planets in identifying and elucidating the properties of dust devils \citep{2016SSRv..203...39M}. In this kind of survey, a sensor is deployed in-situ and records a pressure time series at a sampling rate $\geq 1$ Hz. Since it is a low-pressure convective vortex, the nearby passage of a dust devil will register as pressure dip discernible against a background ambient (but not necessarily constant) pressure. Figure \ref{fig:conditioning_detection_b_inset} adapted from \citep{Jackson_2015} shows a time-series with a typical dust devil signal. 
  
However, single-sensor barometer surveys also suffer from important biases. Foremost among these biases is the fact that a pressure dip does not necessarily correspond to a dust-lofting vortex. Indeed, recent studies \citep{2014DPS....4630006S} suggest pressure dips are often unaccompanied by lofted dust, likely because the attendant wind velocities are not sufficient to lift dust. The problem of identifying dustless vortices can be mitigated if barometers are deployed alongside solar cells, which can register obscuration by dust as long as the dust-lifting devil passes between the sensor and the Sun \citep{Lorenz_2015}.

Another key bias is the ``miss distance'' effect: a fixed barometric sensor is very likely to pass through a dust devil some distance from its center. Since the pressure perturbation falls off with distance, the deepest point in the observed pressure profile will almost always be less than the pressure dip at the devil's center. The observed shape of the profile will be distorted as well. Additional biases can also influence the inferred statistical properties: noise in the pressure time series from a barometer may make more difficult detection of a dust devils with smaller pressure perturbations, depending on the exact detection scheme. 

However, simple geometric considerations can mitigate the influence of the miss distance effect, allowing single-barometer surveys to be statistically corrected. Such statistical considerations are key for understanding the population-averaged influence of dust devils \citep{2016SSRv..203..277L}. In this study, we present a model for correcting the miss distance effect. This study is motivated by \citet{Lorenz_2014}, but, where that study used a numerical simulation to investigate biases in the recovered population of dust devil properties, we employ an analytic framework that provides more direct insight into the problem. 

In Section \ref{sec:Formulating_the_statistical_model}, we develop our statistical model and describe how the geometry of encounter both biases and distorts the recovered parameters. In Section \ref{sec:application_to_observational_data}, we conduct a preliminary application of our model to data from a dust devil survey, and in Section \ref{sec:discussion_and_conclusions}, we discuss some of the limitations of our model, ways to improve it, and future work.

\section{Formulating the Statistical Model}
\label{sec:Formulating_the_statistical_model}

To develop a model for the recovery biases and signal distortions induced by the miss distance effect, we will make the following assumptions:

\begin{enumerate}

\item Each dust devil pressure profile has a well-defined, static radial profile, which follows a Lorentzian: 
\begin{equation}
\label{eqn:lorentzian_profile}
P(r) = \frac{P_{\rm act}}{1 + \left( 2r/\Gamma_{\rm act} \right)^2 },
\end{equation}
where $r$ is the distance from the devil center, $P_{\rm act}$ the actual pressure depth at the devil's center, and $\Gamma_{\rm act}$ the profile full-width at half-max. Alternative profiles have been suggested, (e.g., Burgers-Rott and Vatistas profiles -- \citealp{Lorenz_2014, 2016SSRv..203..209K}), but using a different profile in our analysis would not qualitatively change our results. For instance, a profile that fell off more slowly with radial distance would less significantly bias the recovered devil population toward larger devils (Section \ref{sec:the_recovery_bias}), but the bias would persist.

\item The dust devil center travels at a translation velocity $\upsilon$, which is constant in magnitude and direction. In reality, a devil's trajectory can be complex, even encountering a sensor multiple times and/or a devil may consist of multiple convective cores, consequently producing complex pressure signals \citep{Lorenz_2013}. The time it takes for a devil to travel a distance equal to its own diameter (i.e., full-width/half-max) $\Gamma_{\rm act}$ is $\tau_{\rm act} = \Gamma_{\rm act}/\upsilon$. By contrast, the pressure signal registered by a devil is observed to have a width in time $\tau_{\rm obs} =  \Gamma_{\rm obs}/\upsilon$, where $\Gamma_{\rm obs}$ is the inferred width of the devil's profile.

\item Dust devil velocities follow a smooth distribution $\rho(\upsilon)$ such that the small fraction of dust devils that have velocities between $\upsilon$ and $\upsilon + d\upsilon$ is  $\rho(\upsilon)\ d\upsilon$. Since pressure time-series register devils as signals in time, converting from the devil profile duration to its width requires accounting for the travel velocity. 

\item A dust devil appears and disappears instantaneously, traveling a distance $\upsilon L$ over its lifetime $L$. As pointed out by \citet{Lorenz_2013}, $L$ seems to depend on dust devil diameter $D$ as $L = 40\ {\rm s}\ \left( D/{\rm m} \right)^{2/3}$, with diameter in meters. We assume $D = \Gamma_{\rm act}$ \citep{Vatistas_1991}. 

\item There are minimum and maximum pressure profile depths that can be recovered by a survey, $P_{\rm th}$ and $P_{\rm max}$, respectively. $P_{\rm th}$ may be set by the requirement that a pressure signal exceeds some minimum threshold set by the noise in the datastream, while basic thermodynamic limitations likely restrict the maximum pressure depth for a devil \citep{Renn__1998}. Likewise, the profile widths must fall between $\Gamma_{\rm th}$ and $\Gamma_{\rm max}$, possibly set by the ambient vorticity field in which a devil is embedded \citep{Renn__2001} and/or the depth of the planetary boundary layer \citep{Fenton_2015}. The two sets of limits may not be related, i.e. devils with $P_{\rm max}$ do not necessarily have widths $\Gamma_{\rm max}$. As it turns out, our results are not sensitive to the precise values for each of these limits.

\item The two-dimensional distribution of $P_{\rm act}$ and $\Gamma_{\rm act}$, $\rho(P_{\rm act}, \Gamma_{\rm act})$, is integrable and differentiable. The same is true for the distributions of observed dust devil parameters, $\rho(P_{\rm obs}, \Gamma_{\rm obs})$. It is also important to note that all the distributions shown here are \emph{differential} distributions and not \emph{cumulative} \citep{Lorenz_2011}.

\item The uncertainties on the profile depth and width estimated for a dust devil are negligible. In \citet{Jackson_2015}, for example, uncertainties on $P_{\rm obs}$ were about an order of magnitude less than the inferred $P_{\rm obs}$ value for a detected devil, with uncertainties on $\Gamma_{\rm obs}$ at least a factor of three smaller. Robust recovery of a devil against noise requires relatively small uncertainties.

\end{enumerate}

\begin{figure}
\includegraphics[width=\textwidth]{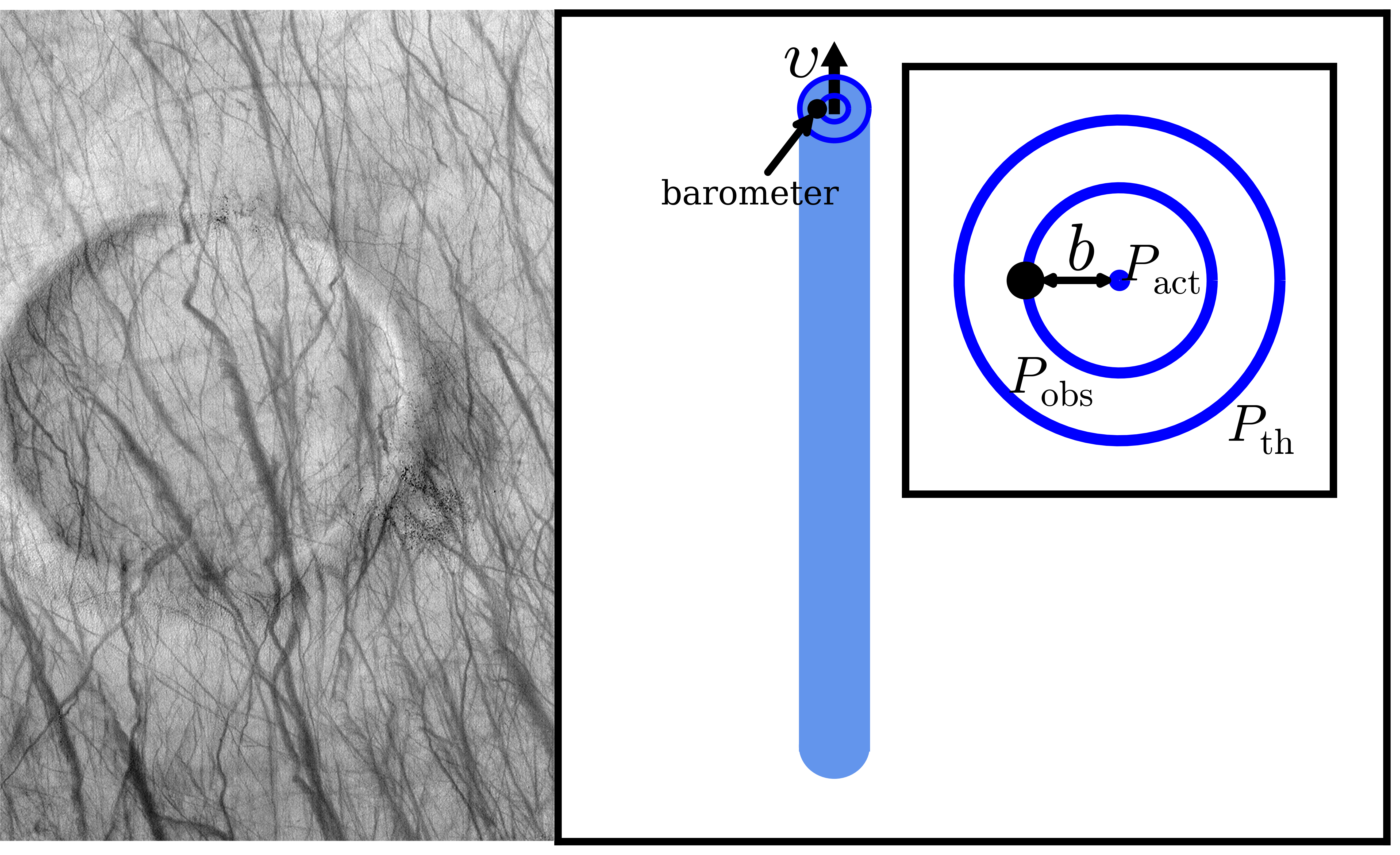}
\caption{(Left) Mars Global Surveyor Mars Orbiter Camera image of crater showing dust devil tracks. The image covers an area 3 km (1.9 mi) wide, is illuminated from the upper left, and is available at \url{http://www.msss.com/mars_images/moc/2003/12/02/2003.12.02.R0901707.gif}. (Right) Geometry of dust devil encounter. The blue circles show pressure contours, the y-axis runs along the devil's translational velocity vector measured relative to the sensor $\mathbf{\upsilon}$, and $b$ is the closest approach distance between the barometer and the devil center. The shaded blue region shows the area of the surface carved out by the traveling $P_{\rm th}$ contour. The inset shows a close-up of the encounter.}
\label{fig:encounter_geometry}
\end{figure}

We can relate the geometry of an encounter directly to the observed profile parameters, and Figure \ref{fig:encounter_geometry} illustrates a typical encounter. As a devil passes the barometer, it will have a closest approach distance $b$, and the radial distance between devil center and sensor $r(t) = \sqrt{b^2 + \left( \upsilon t \right)^2}$, where time $t$ runs from negative to positive values. The fact that $b$ is usually greater than zero biases the devils that are detected and the way in which their pressure signals register. First, we discuss how to convert the distribution of observed dust devil durations to inferred diameters. Next, we formulate the recovery biases and signal distortions resulting from the miss distance effect. 

\subsection{Converting Durations to a Distribution of Diameters}
\label{sec:converting_durations_to_a_distribution_of_diameters}
Very often, the velocity with which a dust devil is traveling as it registers a pressure signal is unknown, which, unfortunately, means the duration of a devil in a time-series $\tau$ cannot be directly converted to a profile width $\Gamma$ (both the observed and actual values are afflicted in the same way with this uncertainty). Via field experiment, \citet{Balme_2012} found that dust devils tend to follow the ambient wind field, and \citet{2016Icar..271..326L} argued that the departure in migration direction from the ambient wind corresponds to an additional translational velocity vector about 2 m/s in magnitude with a random azimuth. Even without a directly measured translation velocity, we can convert statistically from $\upsilon$ and $\tau$ values to $\Gamma$ values.

If we have a collection of discretely measured dust devil durations $\tau_{\rm i}$ from a time-series analysis and an expected distribution of wind speeds, we can ask what is the probability that the ${\rm i}$th dust devil had a $\Gamma$ value between $\Gamma$ and $\Gamma + d\Gamma$. Elementary statistics indicates that this distribution $\rho(\Gamma)$ is
\begin{equation}
\label{eqn:convert_tau-i_and_p-upsilon_to_p-Gamma}
\rho(\Gamma) = \left( \frac{d\upsilon}{d\Gamma} \right)_{\rm i} \left( \frac{dN}{d\upsilon}\right) = \tau_{\rm i}^{-1} \rho(\upsilon = \Gamma/\tau_{\rm i}),
\end{equation}
where $\rho(\upsilon)$ is the probability density for the wind speed.

\begin{figure}
\includegraphics[width=\textwidth]{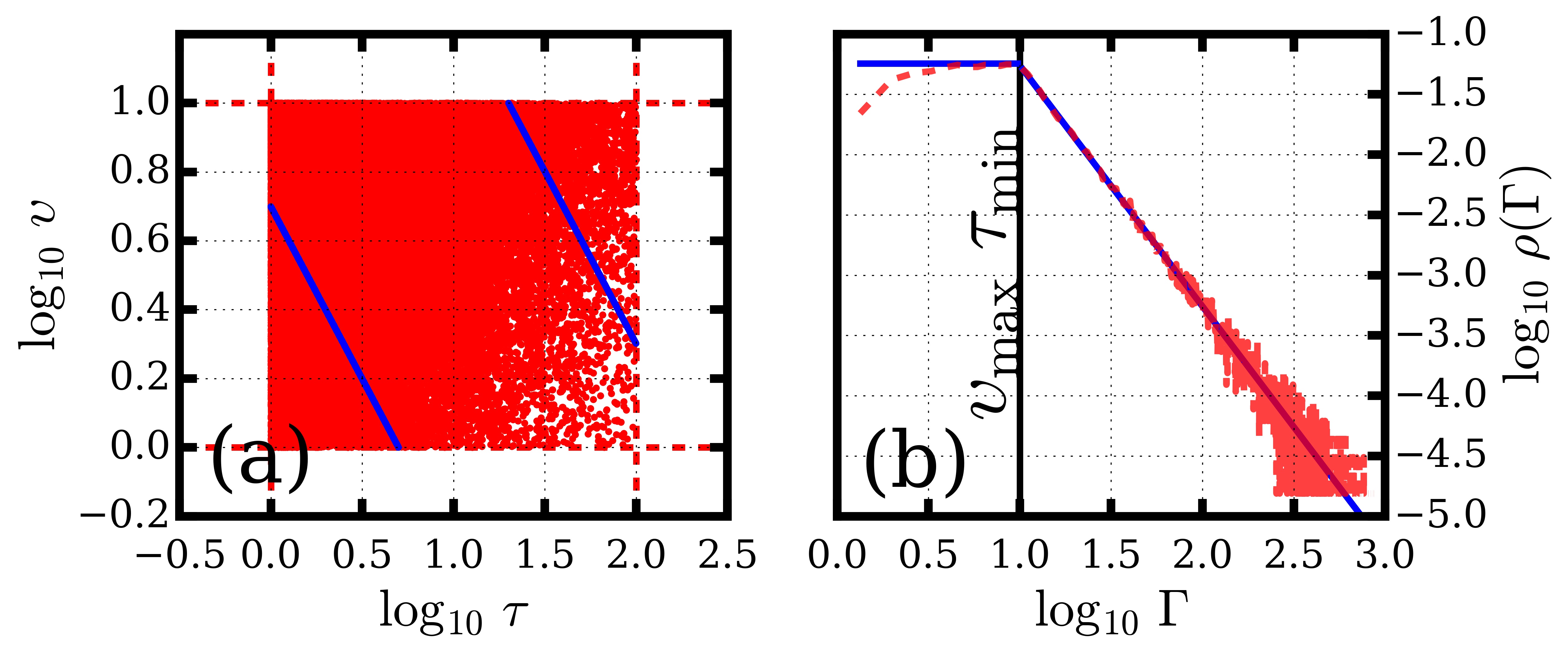}
\caption{(a) A distribution of $\upsilon$- (uniform) and $\tau$ values ($\sim \tau^{-2}$), shown as red dots, along with the integration tracks for calculating $\rho(\Gamma)$, shown as blue curves. (b) The resulting $\rho(\Gamma)$. The dashed, red line shows the direct numerical conversion of the $\upsilon$-$\tau$ distributions from (a), while the solid, blue lines show the analytic prediction. The discrepancies between the two curves at the left and right ends result from binning and sampling distortions.}
\label{fig:converting_rho-tau_to_rho-gamma}
\end{figure}

We can also convert the joint distribution for $\upsilon$ and $\tau$ to a distribution of $\Gamma$ values:
\begin{equation}
\label{eqn:convert_rho-tau-upsilon_to_rho-Gamma}
\rho(\Gamma) = \frac{d}{d\Gamma}\ \int_{D_{\Gamma}} \rho(\upsilon, \tau)\ d\upsilon\ d\tau,
\end{equation}
where $D_{\Gamma}$ is the region in $(\upsilon, \tau)$-space illustrated in Figure \ref{fig:converting_rho-tau_to_rho-gamma} (a). 

From the figure, we can see that when $\Gamma \le \upsilon_{\rm max}\ \tau_{\rm min}$, the integration limits are $\tau \in \left[\tau_{\rm min},\ \mathrm{max}\left( \tau_{\rm max},\ \Gamma/\upsilon_{\rm min} \right)\right]$ and $\upsilon \in \left[ \upsilon_{\rm min},\ \Gamma/\tau \right]$. When $\Gamma > \upsilon_{\rm max}\ \tau_{\rm min}$,\ $\tau \in \left[\Gamma/\upsilon_{\rm max},\ \mathrm{max}\left( \tau_{\rm max}, \Gamma/\upsilon_{\rm min} \right)\right]$ and $\upsilon \in \left[ \upsilon_{\rm min}, \Gamma/\tau \right]$.

In Section \ref{sec:application_to_observational_data}, we calculate this integral numerically using real data, but as an example, consider a simple differential distribution:
\begin{equation}
\rho(\upsilon, \tau) = k \tau^{-\alpha}
\end{equation}
for which velocities are uniformly distributed. The resulting distribution of $\Gamma$ values has a complicated, piece-wise form described in the Appendix and is shown in Figure \ref{fig:converting_rho-tau_to_rho-gamma} (b) for $\alpha = 2$.

\subsection{The Signal Distortion}
\label{sec:the_signal_distortion}
The deepest point observed in the pressure profile $P_{\rm obs}$ is given by 
\begin{equation}\label{eqn:P_obs}
P_{\rm obs} = \frac{P_{\rm act}}{1 + \left( 2 b/\Gamma_{\rm act} \right)^2}.
\end{equation}
Clearly, unless $b = 0$, $P_{\rm obs} < P_{\rm act}$. Likewise, non-central encounters will distort the profile full-width/half-max, giving a full-width/half-max $\Gamma_{\rm obs}$. 

The observed pressure signal drops to half its value at a time $t = \pm \frac{1}{2} \tau_{\rm obs} = \pm \frac{1}{2} \Gamma_{\rm obs}/\upsilon$ by definition. At these times, the center of the devil is a radial distance from the barometer $r(t = \pm \tau_{\rm obs}/2) = \sqrt{b^2 + \left( \frac{1}{2} \Gamma_{\rm obs} \right)^2}$ and 
\begin{equation}
P(r) = \frac{1}{2} P_{\rm obs} = \frac{1}{2} \frac{P_{\rm act}}{1 + \left( 2 b /\Gamma_{\rm act} \right)^2} = \frac{P_{\rm act}}{1 + \left( 2r(\pm \tau_{\rm obs}/2)/\Gamma_{\rm act} \right)^2 }.
\end{equation}

\begin{figure}
\includegraphics[width=\textwidth]{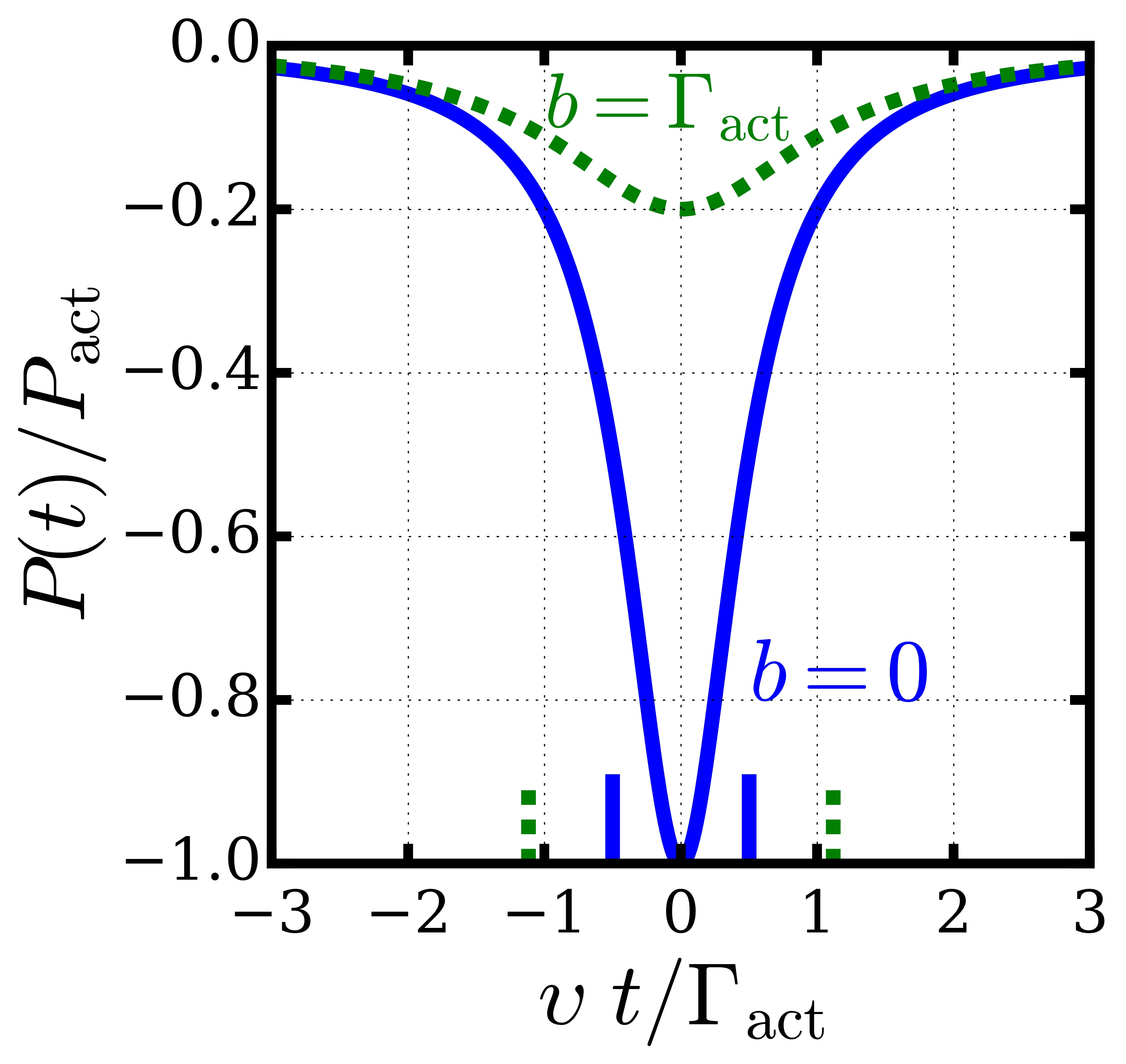}
\caption{Dust devil profiles for $b = 0$ and $b = \Gamma_{\rm act}$, i.e. for a closest approach equal to the profile's diameter. The vertical solid, blue and dashed, green lines at the bottom of the plot show the respective $\Gamma_{\rm obs}$ values.}
\label{fig:compare_profiles}
\end{figure}

Solving for $\Gamma_{\rm obs}$ gives 
\begin{equation}
\label{eqn:Gamma_obs}
\Gamma_{\rm obs}^2 = \Gamma_{\rm act}^2 + \left( 2b \right)^2. 
\end{equation}
Figure \ref{fig:compare_profiles} shows how a non-central encounter modifies the observed pressure profile. $\Gamma_{\rm obs}$ for the green curve with $b = \Gamma_{\rm act}$ is $\sqrt{1 + 2^2} \approx 2.2$ times larger than that for the blue curve with $b = 0$. We call this modification the signal distortion. Combining the Equations \ref{eqn:P_obs} and \ref{eqn:Gamma_obs} gives the following useful expression:
\begin{equation}
\label{eqn:P_obs_Gamma_obs}
P_{\rm obs}\ \Gamma_{\rm obs}^2 = P_{\rm act}\ \Gamma_{\rm act}^2. 
\end{equation}

We can solve Equation \ref{eqn:P_obs} for $b$:
\begin{equation}\label{eqn:b}
b = \left( \frac{\Gamma_{\rm act}}{2} \right) \sqrt{ \frac{P_{\rm act} - P_{\rm obs}}{P_{\rm obs}}}.
\end{equation}
A single barometer at a fixed location can sense a dust devil only over a certain area, spanning a maximum radial distance $b_{\rm max}$, beyond which devils will produce pressure signals smaller than the detection threshold, $P_{\rm obs} < P_{\rm th}$:
\begin{equation}
b_{\rm max} = \left( \Gamma_{\rm act}/2 \right) \sqrt{ \frac{ P_{\rm act} - P_{\rm th} }{P_{\rm th}}}.
\label{eqn:b_max}
\end{equation}

We can use the encounter geometry to model the statistical probability for $P_{\rm obs}$ and $\Gamma_{\rm obs}$ to fall within a certain range of values, given a distribution of $P_{\rm act}$ and $\Gamma_{\rm act}$ values. The probability density for passing between $b$ and $b + db$ of a devil is $dp(b) = 2 b\ db / b_{\rm max}^2 $ for $b \le b_{\rm max}$. This expression allows us to calculate the average miss distance for a given dust devil: 
\begin{equation}
\langle b/\Gamma_{\rm act} \rangle = \int b/\Gamma_{\rm act}\ dp = \frac{2}{3}\ b_{\rm max}/\Gamma_{\rm act},
\label{eqn:average_b}
\end{equation}
which is $\approx 1/3 \sqrt{P_{\rm act}/P_{\rm th}}$ for $P_{\rm act} \gg P_{\rm th}$. If, for example, $P_{\rm act} \approx 10\ P_{\rm th}$, $\langle b \rangle \approx \Gamma_{\rm act}$, meaning that, on average, $P_{\rm obs} \approx P_{\rm act}/5$ and $\Gamma_{\rm obs} \approx 5\ \Gamma_{\rm act}$.

\subsection{The Recovery Bias}
\label{sec:the_recovery_bias}
\begin{figure}
\includegraphics[width=\textwidth]{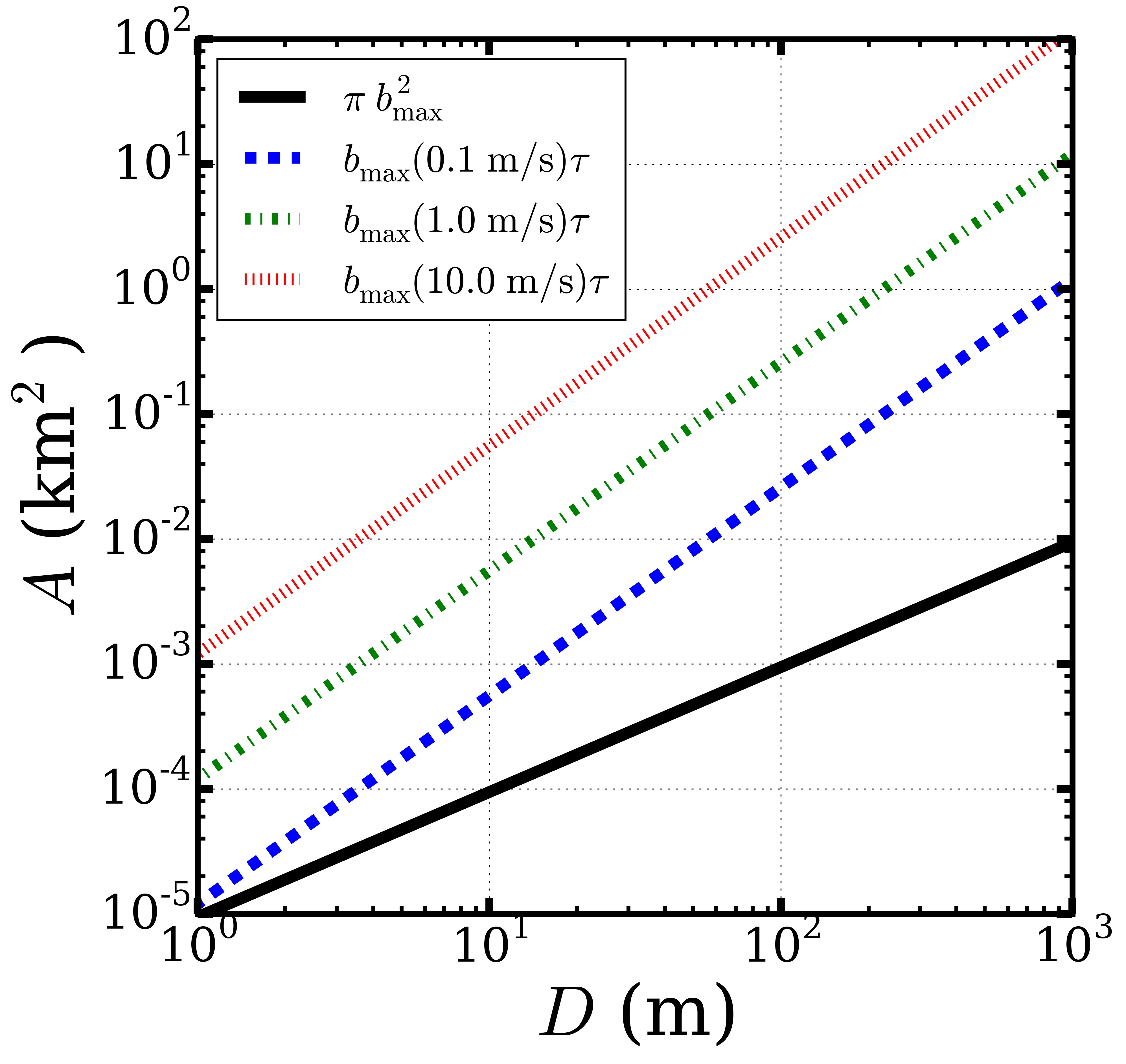}
\caption{The two area terms from Equation \ref{eqn:dust_devil_area} for dust devils with a range of diameters $D$ traveling with a range of velocities, 0.1, 1, and 10 m/s.}
\label{fig:relative_areas}
\end{figure}

As it travels on the surface of the observational arena, a dust devil's pressure contour $P_{\rm th}$ carves out a long, narrow area $A(P_{\rm act}, \Gamma_{\rm act})$. If a barometer lies within that area, the devil will be detected, in principle. $A$ is given by 
\begin{eqnarray} 
\label{eqn:dust_devil_area}
A & = & \pi b_{\rm max}^2 + \upsilon L b_{\rm max}\nonumber \\ 
  & = & \left( \Gamma_{\rm act}/2 \right) \sqrt{ \frac{P_{\rm act} - P_{\rm th}}{P_{\rm th}} } \left[ \pi \left( \Gamma_{\rm act}/2 \right) \sqrt{ \frac{P_{\rm act} - P_{\rm th}}{P_{\rm th}} } + \upsilon L \right],
\end{eqnarray}
The probability to recover a devil is proportional to this total track area. Thus devils with deeper and wider pressure profiles are more likely to be recovered. Using the lifetime scaling from \citet{Lorenz_2014}, Figure \ref{fig:relative_areas} shows that the second term dominates over the first term for all but the smallest, slowest dust devils, so, for simplicity, we will neglect the first term, giving 
\begin{equation}
A \approx \left( \Gamma_{\rm act}/2 \right) \sqrt{ \frac{P_{\rm act} - P_{\rm th}}{P_{\rm th}} } \upsilon L.
\end{equation}

The fact that larger, faster dust devils cover more area means that they are more likely to be recovered. We can quantify the recovery probability $f$ by taking the ratio of track areas for a given dust devil to the largest area for a dust devil, $A_{\rm max}$:
\begin{equation}
\label{eqn:recovery_bias}
f = \frac{A(\Gamma_{\rm act}, P_{\rm act})}{A_{\rm max}} = A_{\rm max}^{-1}\ \Gamma_{\rm act} \sqrt{\frac{P_{\rm act} - P_{\rm th}}{P_{\rm th}}} \upsilon\ L.
\end{equation}
The devil with the deepest profile need not also have the widest profile or the largest translation velocity. \citet{Renn__2001} argue that the diameter of a vortex is set, in part, by the local vorticity field, while \citet{Balme_2012}, from their field studies, find no correlation between diameter and translation velocity from their field work. In quantifying the recovery probability $f(\Gamma_{\rm act}, P_{\rm act})$, it is only important that we apply a uniform normalizing factor to the whole population, so we will take $A_{\rm max} = \Gamma_{\rm max} \sqrt{\left( P_{\rm max} - P_{\rm th} \right)/P_{\rm th}}\ \upsilon L_{\rm max}$, assuming no correlation between translation velocity and diameter or pressure, which gives
\begin{equation}
f = \bigg[ \Gamma_{\rm max}^{-1} L_{\rm max}^{-1} \left( P_{\rm max} - P_{\rm th} \right)^{-1/2} \bigg]\ \Gamma_{\rm act} L \left( P_{\rm act} - P_{\rm th} \right)^{1/2}
\label{eqn:final_f_equation}
\end{equation}

Taking the distribution of observed devils as $\rho(\Gamma_{\rm act}, P_{\rm act})$, the product $f(\Gamma_{\rm act}, P_{\rm act}) \times \rho(\Gamma_{\rm act}, P_{\rm act})$ would represent the population of devils that are detected but not how the recovered population would actually look.

\subsection{Converting Between the Observed and Actual Parameter Distributions}
\begin{figure}
\includegraphics[width=\textwidth]{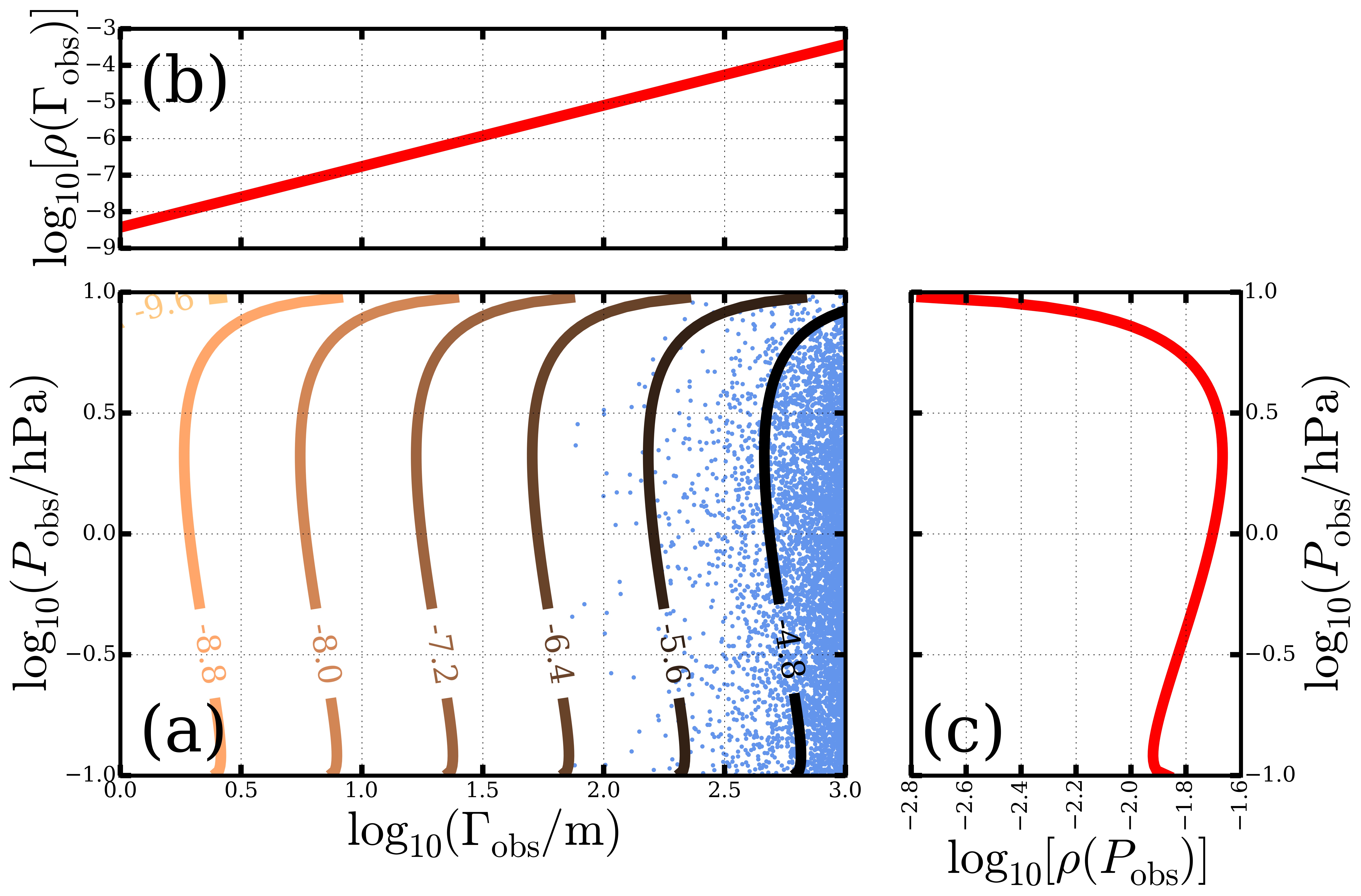}
\caption{(a) Contours of number density for observed dust devil parameters, $\log_{10}\ \rho(P_{\rm obs}, \Gamma_{\rm obs})$, assuming a uniform distribution of underlying values $\rho(P_{\rm act}, \Gamma_{\rm act}) = \left( P_{\rm max} - P_{\rm th} \right)^{-1}\ \left( \Gamma_{\rm max} - \Gamma_{\rm th} \right)^{-1}$. The blue dots show a simulated dust devil survey. (b) Density of $\Gamma_{\rm obs}$ integrated over $P_{\rm obs}$. (c) Density of $P_{\rm obs}$ integrated over $\Gamma_{\rm obs}$.}
\label{fig:uniform_actual_distribution_to_observed_distribution}
\end{figure}

 Consider a distribution of observed values $\rho(\Gamma_{\rm obs}, P_{\rm obs}) = d^2N/d\Gamma_{\rm obs}\ dP_{\rm obs}$. The small number of devils $dN = f\ \rho(\Gamma_{\rm obs}, P_{\rm obs})\ d\Gamma_{\rm act}\ dP_{\rm act}$ contributing are those that had closest approach distances between $b$ and $b + db$ of the detector. Thus, we can convert $\rho(\Gamma_{\rm act}, P_{\rm act})$ to $\rho(\Gamma_{\rm obs}, P_{\rm obs})$ by integrating $\rho(\Gamma_{\rm act}, P_{\rm act})$ over $b$ and accounting for the bias and distortion effects:
\begin{equation}
\label{eqn:convert_from_actual_to_observed_density}
\rho({\rm obs}) = \int_{b^\prime = 0}^{b({\rm obs})} f\ \rho({\rm act}(b^\prime))\ \frac{2b^\prime\ db^\prime}{b_{\rm max}^2},
\end{equation}
where $b({\rm obs})$ $= \left(\Gamma_{\rm obs}/2\right) \left[ \left(P_{\rm max} - P_{\rm obs}\right)/P_{\rm max} \right]^{1/2}$, i.e. the maximum radial distance at which the actual distribution could contribute to the observed distribution. Figure \ref{fig:uniform_actual_distribution_to_observed_distribution} shows the result for a uniform distribution of actual values, $\rho({\rm act}) = \left( P_{\rm max} - P_{\rm th} \right)^{-1}\ \left( \Gamma_{\rm max} - \Gamma_{\rm th} \right)^{-1}$ and compares it to the simulated results of an observational survey (blue circles). 

Ultimately, we are interested in converting the density of observed parameters to the density of actual parameters. We describe the details of our approach in the Appendix, but the basic idea is that $\rho(\Gamma_{\rm obs}, P_{\rm obs})$ represents an integral sampling $\rho(\Gamma_{\rm act}, P_{\rm act})$ along a locus of points $(\Gamma_{\rm act}, P_{\rm act})$, going from a point representing $b = 0$ up to a point representing the maximum radial distance. Thus, we can differentiate $\rho(\Gamma_{\rm obs}, P_{\rm obs})$ to determine $\rho(\Gamma_{\rm act}, P_{\rm act})$:
\begin{equation}
\label{eqn:the_big_equation}
\rho(\Gamma_{\rm act}, P_{\rm act}) =  k \Gamma_{\rm act}^{-5/3}\ \left( P_{\rm act} - P_{\rm th} \right)^{-1/2} \left( P_{\rm obs} \frac{\partial \rho(\Gamma_{\rm obs}, P_{\rm obs})}{\partial P_{\rm obs}} - \left( \frac{\Gamma_{\rm obs}}{2} \right) \frac{\partial \rho(\Gamma_{\rm obs}, P_{\rm obs})}{\partial \Gamma_{\rm obs}} \right)_{\rm obs \rightarrow act},
\end{equation}
where $k$ is constant and ${\rm obs \rightarrow act}$ indicates that ${\rm obs}$ quantities should be replaced with ${\rm act}$ quantities after the derivatives are taken.

In the next section, we apply Equation \ref{eqn:the_big_equation} to a dataset from a real survey, but as an example, consider the simple differential distribution $\rho(\Gamma_{\rm obs}, P_{\rm obs}) = \alpha\ P_{\rm obs}^{-2}$. Applying Equation \ref{eqn:the_big_equation} gives a distribution of actual parameters: 
\begin{equation}
\rho(\Gamma_{\rm act}, P_{\rm act}) = k^\prime \Gamma_{\rm act}^{-5/3} \left( P_{\rm act} - P_{\rm th} \right)^{-1/2} P_{\rm act}^{-2}.\label{eqn:dist_actual_parameters}
\end{equation}
Note that, for this example, $\partial \rho({\rm obs})/\partial P_{\rm obs} < 0$. In such a case, the signs on the partial derivatives should be flipped since the limits on the integral for Equation \ref{eqn:convert_from_actual_to_observed_density} would be flipped.

Equation \ref{eqn:dist_actual_parameters} increases without limit as $P_{\rm act} \rightarrow P_{\rm th}$ because such shallow dips are only observed for statistically impossible central encounters ($b = 0$). If we assume $P_{\rm th} \ll P_{\rm act}$ for any observed values, then we have:
\begin{equation}
\rho(\Gamma_{\rm act}, P_{\rm act}) \approx k^\prime \Gamma_{\rm act}^{-5/3} P_{\rm act}^{-5/2}\label{eqn:simplified_dist_actual_parameters}.
\end{equation}

Using a direct numerical simulation of a dust devil barometric survey, \citet{Lorenz_2014} found that an observed distribution $\rho(P_{\rm obs}) \propto P_{\rm obs}^{-2}$ required an actual distribution approximately $\rho(P_{\rm act}) \propto P_{\rm act}^{-2.8}$, in line with our results here. 

\section{Application to Observational Data}
\label{sec:application_to_observational_data}
\begin{figure}
\includegraphics[width=\textwidth]{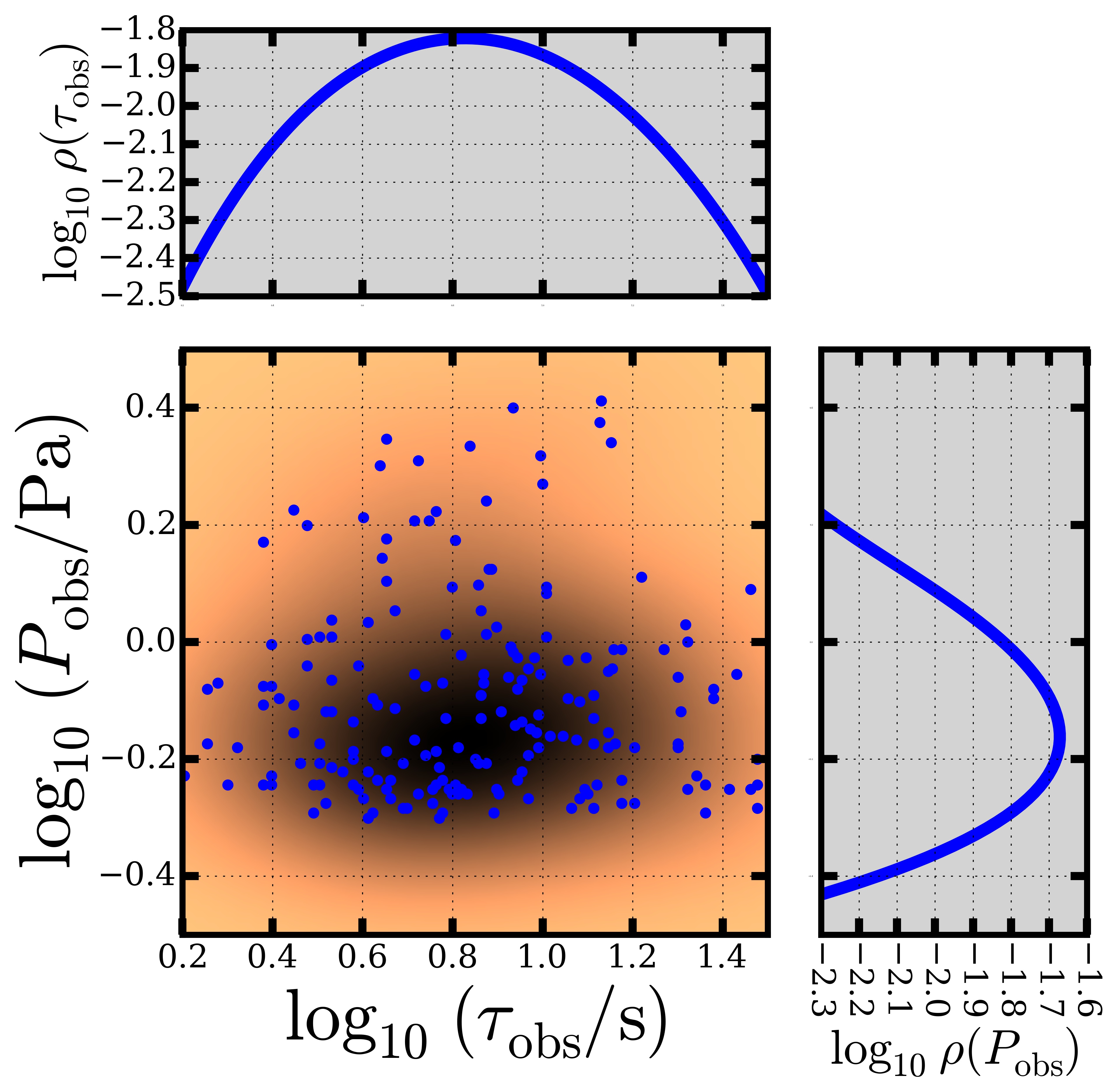}
\caption{The bottom left panel shows a scatter plot (blue dots) of Martian dust devil encounters observed by the meteorological instruments on the Pheonix Lander and reported in Table 1 from \citet{Ellehoj_2010}. The copper-to-black background shows the corresponding density estimated using a Gaussian kernel density estimator, with darker colors indicating higher densities. The blue lines in the top and right panels show the marginalized, normalized distributions of $\tau_{\rm obs}$ and $P_{\rm obs}$, respectively.}
\label{fig:Ellehoj_data_obs_dist}
\end{figure}

In this section, we apply our statistical formulation to results from a real survey, \citet{Ellehoj_2010}, a study using time-series from the Phoenix Lander \citep{Smith_2008}. \citet{Ellehoj_2010} analyzed 151 sols worth of data, including pressures, temperatures, wind speeds, and images. To identify dust devil passages in the barometric data, they compared the average pressure in a 20-s window to the average pressure in 20-s windows to either side of the former window. Average pressures in the middle window that were different by more than 0.1 Pa from the average on either side were identified as possible dust devil passages. Then for every pressure event found, the authors analyzed the surrounding pressure and temperature values, and non-significant and false events, e.g., from data transfer gaps, were removed by hand (the precise criteria used to exclude an event are not given). In this way, \citet{Ellehoj_2010} identified 197 vortices with a pressure drops larger than 0.5 Pa = $10^{-0.3}$ Pa, which we will take as $P_{\rm th}$ for this dataset. Figure \ref{fig:Ellehoj_data_obs_dist} shows a scatter plot of their reported detections. The shaded background is calculated using a Gaussian kernel density estimator and a bandwidth of 0.75, analogous to histogram bin widths \citep{SCOTT_1979}. To determine the distribution of $\tau_{\rm obs}$, we integrated the 2-D density over $P_{\rm obs}$ (i.e., marginalized it), and we show the resulting distribution in the top panel (the right panel shows the analogous result for $P_{\rm obs}$). 

\begin{figure}
\includegraphics[width=\textwidth]{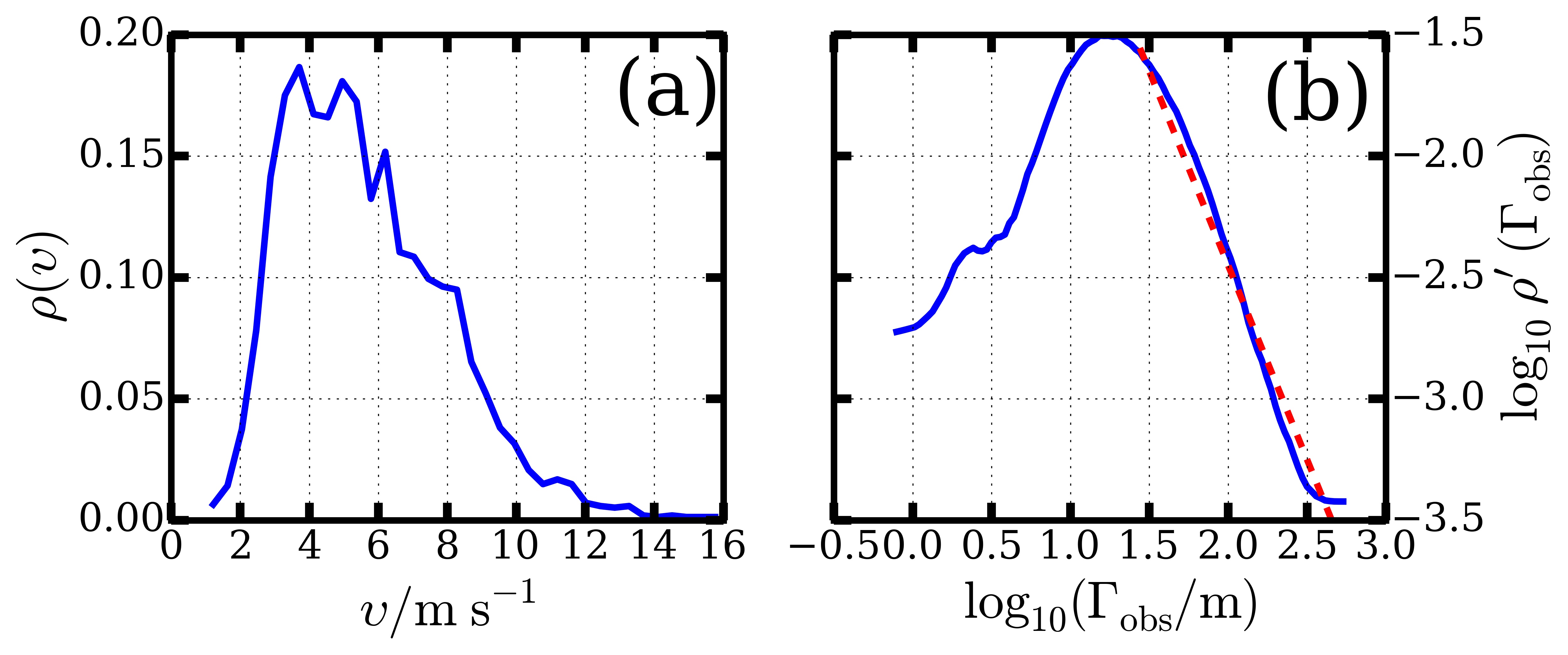}
\caption{(a) Distribution of windspeeds measured by the Phoenix mission \citep{2010JGRE..115.0E18H}. (b) Inferred normalized distribution of dust devil widths using data from \citet{Ellehoj_2010} (solid, blue curve), along with power-law fit having index about -1.6 (dashed, red curve).}
\label{fig:rho-Gammaobs_from_Ellehoj}
\end{figure}

To convert the $\tau_{\rm obs}$ values to a distribution of $\Gamma_{\rm obs}$, we used Equation \ref{eqn:convert_tau-i_and_p-upsilon_to_p-Gamma} and the wind speeds measured during sols 91 through 150 from the Phoenix mission \citep{2010JGRE..115.0E18H}\footnote{Available via NASA's Planetary Data System - \url{http://pds-atmospheres.nmsu.edu/pdsd/archive/data/phx-m-tt-5-wind-vel-dir-v10/phxwnd_0001/DATA/TELLTALE_91_151.TAB}} (Figure \ref{fig:rho-Gammaobs_from_Ellehoj} (a)). These wind speeds were measured primarily between 0700 and 1900 local mean solar time, which overlaps with times when dust devils are most common, but they are not all measured contemporaneously to dust devil activity. In other words, the distribution in Figure \ref{fig:rho-Gammaobs_from_Ellehoj} (a) is broadly representative of daytime wind speeds seen during the Phoenix mission but not necessarily of the wind speeds during dust devil encounters.

Figure \ref{fig:rho-Gammaobs_from_Ellehoj} (b) shows the resulting $\Gamma_{\rm obs}$ distribution. Consistent with estimates of the distributions of dust devil widths \citep[e.g.][]{2016Icar..266..315R}, wider devils are less common than narrow devils, and Figure \ref{fig:rho-Gammaobs_from_Ellehoj} (b) shows a power-law model for the differential distribution with index -1.6 provides a reasonable fit. The distribution also declines for dust devils narrower than $10^{1.25}\ {\rm m}\ \approx 18$ m, seemingly at odds with the results of \citet{2016Icar..266..315R}, which conducted a visual survey of dust devil tracks and reported a population dominated by devils with a diameters $< 10$ m. However, the apparent decline on the narrow-end of the distribution is likely the result of selection bias in \citet{Ellehoj_2010}. Given the typical wind velocity of 5 m/s measured at the Phoenix site, a devil with a width $\sim10$ m would produce a signal with only a $\sim$2-s duration. The search technique from \citet{Ellehoj_2010} requires a devil to induce a deep or lengthy ($\sim$20-s) pressure perturbation in order to register a detection, thus filtering out the smallest (and possibly most common) dust devils. Likewise, pressure perturbations spanning much more than 20 s (presumably, the widest and/or most slowly moving devils) would probably also be filtered out, which may explain why the power-law fit in Figure \ref{fig:rho-Gammaobs_from_Ellehoj} (b) is shallower than in \citet{2016Icar..266..315R}.

\begin{figure}
\includegraphics[width=\textwidth]{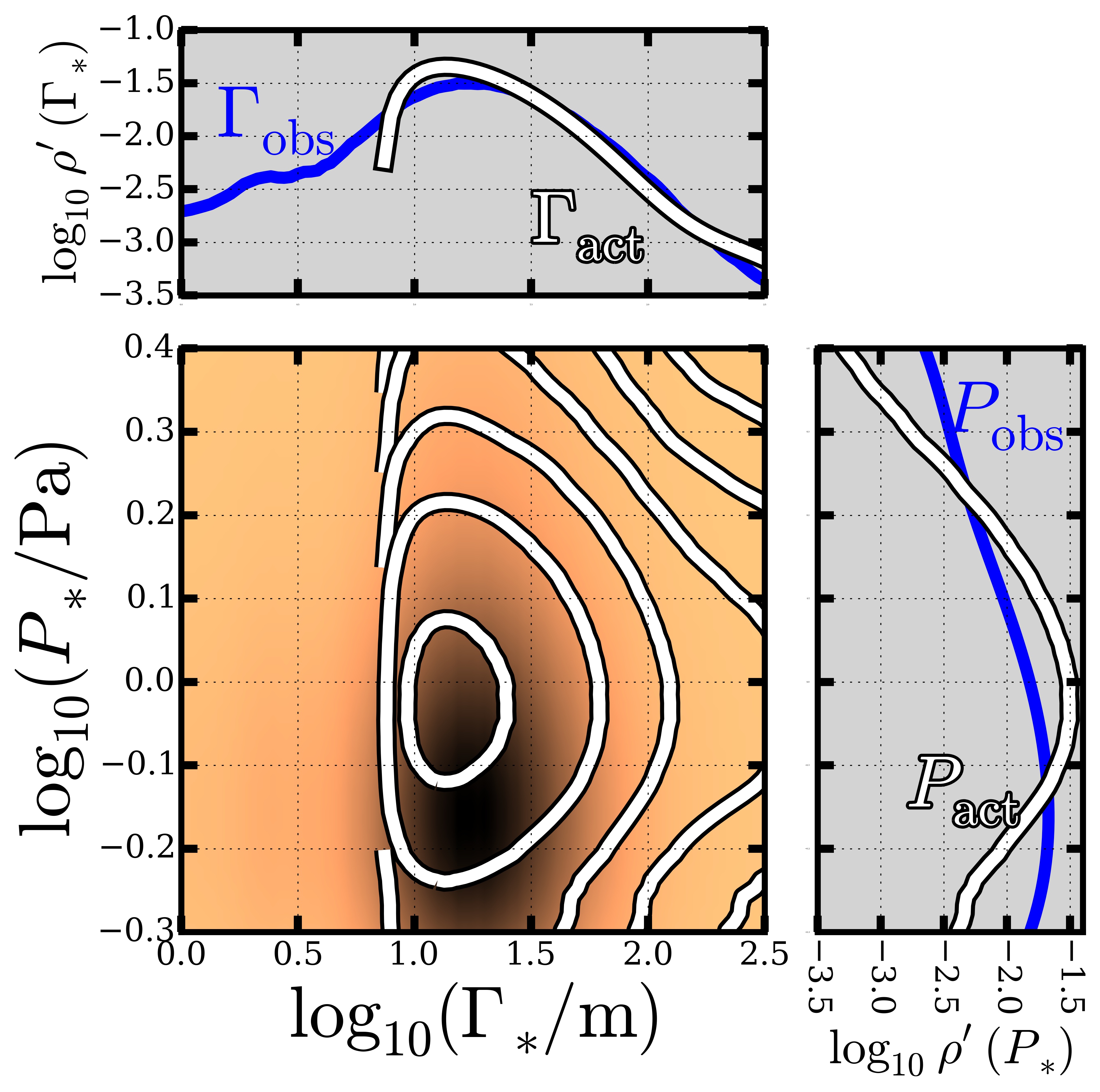}
\caption{The conversion between the joint distribution for observed parameters $\rho(\Gamma_{\rm obs}, P_{\rm obs})$ to that for actual parameters $\rho(\Gamma_{\rm act}, P_{\rm act})$. The copper-to-black shading in the bottom, left panel shows $\rho(\Gamma_{\rm obs}, P_{\rm obs})$, with darker colors indicating higher densities, while the white contours show $\rho(\Gamma_{\rm act}, P_{\rm act})$. The blue lines in the top and right panels show the marginalized, normalized distributions for each of the observed and actual parameters.}
\label{fig:Ellehoj_data_obs_to_act_dist}
\end{figure}

Finally, we can use Equation \ref{eqn:the_big_equation} to convert the joint distribution of observed parameters $\rho(\Gamma_{\rm obs}, P_{\rm obs})$ to the distribution of actual parameters $\rho(\Gamma_{\rm act}, P_{\rm act})$, and the result is shown in Figure \ref{fig:Ellehoj_data_obs_to_act_dist}. The differential distribution for $\Gamma_{\rm act}$ is slightly shallower than that for $\Gamma_{\rm obs}$ and is best fit using a power-law index of -1.5, as compared to -1.6 for $\Gamma_{\rm obs}$, and the mode of the distribution shifts from 18 m for $\Gamma_{\rm obs}$ to 13 m for $\Gamma_{\rm act}$. The distribution of $P_{\rm act}$ shifts toward larger pressures compared to the distribution for $P_{\rm obs}$, with the mode going from 0.7 to 0.9 Pa. As for the distribution of $\Gamma_{\rm obs}$, selection effects probably suppress the distribution of $P_{\rm obs}$ for smallest values since devils with smaller pressure dips have a smaller signal-to-noise ratio. For values in the differential distribution of $P_{\rm obs}$ larger than 0.7 Pa, a power-law model with index about -2 provides a best-fit, while for the $P_{\rm act}$ distribution, an index of -3.5 provides a best-fit. Keep in mind that this distribution of $P_{\rm act}$ comes from marginalizing over $\Gamma_{\rm act}$ the result of applying Equation \ref{eqn:the_big_equation}, which is why it is different from the distribution described at the end of the last section.

Since Equation \ref{eqn:the_big_equation} involves a difference between derivatives, it is possible for the inferred distribution of actual values to become negative, even though negative values are not physically meaningful. As we have argued here, the drops in the distributions at small $\Gamma_{\rm obs}$ and $P_{\rm obs}$ probably arise from selection effects and are not physical. In order to avoid retrieving negative values for our distributions, we have focused on evaluating Equation \ref{eqn:the_big_equation} in regions where the distribution remains positive, where selection effects are probably less important. 

By averaging Equation \ref{eqn:average_b} over the population, we can estimate the population-weighted average miss distance:
\begin{equation}
\langle \langle b \rangle \rangle = \frac{2}{3}\ \int b_{\rm max}/\Gamma_{\rm act}\ \rho({\rm act})\ d\Gamma_{\rm act}\ dP_{\rm act} \approx 16\ {\rm m},
\end{equation}
or roughly one full-width/half-max. For this calculation, we only integrated over the population where $\Gamma_{\rm act} \geq$ 13 m and $P_{\rm act} \geq $ 0.7 Pa to mitigate the effects of the detection biases, which likely suppress the observed number of narrow and shallow dust devil signals.

\section{Discussion and Conclusions}
\label{sec:discussion_and_conclusions}
Our formulation here provides a starting place for relating the population statistics of dust devils as recovered by single-barometer surveys to their physical structures, provided some measure of their translational velocities. Understanding these relationships is critical for understanding the atmospheric influence of devils on both planets since it depends so sensitively on both the devils' statistical and physical properties. As noted in \citet{Jackson_2015} and \citet{Lorenz_2014}, in estimating the total flux of dust injected into the martian atmosphere, it is important to consider the population-weighted flux and not the flux from the average dust devil, which \citet{Jackson_2015} showed can amount to several orders of magnitude increase in the dust flux. The bias/distortion corrections to the populations discussed here increases the flux by at least 20\% over that. Accounting for biases in the time-series analysis could increase the flux even more since, as discussed here, the scheme used in \citet{2010JGRE..115.0E18H} likely suppresses the recovery of the larger/slow-moving vortices, which may lift the most dust, and smaller/quickly moving dust devils, which are the most numerous.

%
%

As discussed in Section \ref{sec:the_recovery_bias}, the miss distance effect biases the recovered population toward the physically widest devils. Because the dynamical processes that form and maintain devils are not well-understood, the relationship between the width of a devil and its other physical properties are not clear. \citet{Fenton_2015} argue that dust devil height is related to the boundary layer depth, while the physical model outlined in \citet{Renn__1998} indicates the profile depth should also scale with boundary layer depth. In any case, the bias definitely plays a role in estimates of the areal density for dust devil occurrence. For example, by assuming a devil profile width of 100 m, \citet{Ellehoj_2010} combine the number of devils recovered from pressure time-series and wind speed data to estimate a local occurrence rate of 1 event per sol per 10 km$^2$. Our analysis here (Figure \ref{fig:rho-Gammaobs_from_Ellehoj}) shows that such wide devils were probably not typical among those detected. Instead, we can adapt their formulation for our estimated typical, actual diameter of 13 m (Figure \ref{fig:Ellehoj_data_obs_to_act_dist}), which gives an occurrence rate ten times higher, 1 event per sol per km$^2$. Investigating the abundance of dust devil tracks in the southern hemisphere in Argyre Planitia and Hellas Basin, \citet{2003JGRE..108.5086B} estimated rates as high as about 0.2 events per sol per km$^2$, suggesting that something like only one in five low-pressure cells lift sufficient dust to leave a visible track. This estimate includes detection biases and a population average should be used instead of typical values, but these results are roughly consistent with terrestrial field studies that estimate 40\% of dust devils lift visible amounts of dust \citep{Lorenz_2015}.

The model for the miss distance effect developed here serves to highlight the many important uncertainties and degeneracies involved in single-barometer dust devil surveys, in particular, the difficulty of disentangling the geometry of an encounter between a devil and a detector from the devil's structure. However, the encounter geometry can be determined if additional measurements are made. For instance, if the tangential velocity profile for a dust devil can be measured simultaneously with the pressure profile, the miss distance can be determined directly by assuming cyclostrophic balance, i.e. $ V_{\rm T}^2 = \Delta P_{\rm act}/\rho $, a pressure profile given by Equation \ref{eqn:lorentzian_profile}, and one of the physically motivated vortex profiles given in \citet{Vatistas_1991} for the tangential windspeed:
\begin{equation}
V(r) = \frac{2 r V_{\rm T}}{1 + \left( 2 r/\Gamma_{\rm act} \right)^2},
\label{eqn:tangential_windspeed}
\end{equation}
where $V_{\rm T}$ is the windspeed at the eyewall of the vortex and $\rho$ is the atmospheric density. With these assumptions, the pressure $P$ and velocity $V$ measured at any point in the profile obey
\begin{eqnarray}
P_{\rm act} & = & \left( 1 - \dfrac{\rho V^2}{4 P} \right)^{-1} P \label{eqn:DeltaP0_from_DeltaP},\ {\rm and}\ \\
V_{\rm T} & = & \left( 1 - \dfrac{\rho V^2}{4 P} \right)^{-1/2} \left( \dfrac{P}{\rho} \right) ^{1/2} \label{eqn:VT_from_V}.
\end{eqnarray}
Unfortunately, many in-situ surveys produce only temperature time-series along with the pressure time-series, but a clear prediction of the temperature profile might allow a determination of the miss distance directly from the pressure and temperature time-series.

An improved understanding of the biases involved in a detection scheme is critical for relating the observed to the underlying population, and a simple way to assess a scheme's detection efficiency is to inject synthetic devil signals (with known parameters) into the real data streams. Then the detection scheme can be applied to recover the synthetic devils and the efficiency of detection assessed across a swath of devil parameters. Such an approach is common in exoplanet transit searches \citep[e.g.][]{Sanchis_Ojeda_2014}, where dips in photometric time series from planetary shadows closely resemble dust devil pressure signals. By injecting synthetic devils into real data, the often complex noise structure in the data is retained and simplifying assumptions (such as stationary white noise) are not required. 

Among important limitations of our model, the translation velocity $\upsilon$ for devils remains a critical uncertainty for relating physical and statistical properties. This limitation points to the need for wind velocity measurements made simultaneously with pressure measurements in order to accurately estimate dust devil widths. In particular, correlations between $\upsilon$ and dust devil properties will skew the recovered parameters in ways not captured here. For example, the devils with the deepest pressure profiles seem to occur preferentially around mid-day local time both on Mars \cite{Ellehoj_2010} and the Earth \cite{Jackson_2015}. If winds at that time of day are preferentially fast or slow, then the profile widths recovered for the deepest devils will be skewed toward smaller or larger values. In addition, some field observations suggest devils with larger diameters may be advected more slowly than their smaller counterparts \cite{Greeley_2010}, which would tend to make their profiles look wider.  

Clear predictions of the distributions of physical parameters for dust devils from high resolution meteorological models would be especially helpful for constraining and directing this work, and some progress in this area has been made. For example, \citet{2005QJRMS.131.1271K} applied a large-eddy simulation of a planetary convective boundary layer to study vortical structures and the influence of ambient conditions on their formation. For the handful of vortices formed in the simulations, there was good qualitative agreement with observation. \citet{2010BoLMe.137..223G} also studied vortex formation on Earth and Mars and noted the role of the boundary layer's depth on vortex scale. Given the stochastic nature of boundary layer dynamics, detailed statistical predictions from such models are needed for comparison to observation. Recently, \citet{2016GeoRL..43.4180N} have systematically explored size-frequency distributions in a large eddy simulation run at different resolutions. 

Likely the best way to study dust devil formation and dynamics in the field is not statistically, but directly via deployment of sensor networks that produce a variety of data streams with high spatial and time resolution. Field work with in-situ sensors has a long history but usually involving single-site deployments \citep[e.g.][]{Sinclair_1973}. In the decades since that study, technological developments in miniaturization and data storage now provide a wealth of robust and inexpensive instrumentation, ideally suited for the long-term field deployment required to study dust devils, without the need for direct human involvement. Recently, \citet{2015AeoRe..19..183L} deployed an array of ten miniature pressure- and sunlight-logging stations at La Jornada Experimental Range in New Mexico, providing a census of vortex and dust-devil activity at this site. The simultaneous measurements resolved horizontal pressure structures for several dust devils, giving entirely independent estimates of vortex size and intensity. 

The rich and growing databases of high-time-resolution meteorological data, both for the Earth and Mars, combined with the wide availability and affordability of robust instrumentation, point to bright future for dust devil studies. The data streaming in from the Mars Science Laboratory Rover Environmental Monitoring Station (REMS) \citep{G_mez_Elvira_2012} may provide new insight into Martian dust devils, and recent studies using imaging data from the Mars Science Laboratory \citep[e.g.][]{2017LPI....48.2952L} have spotted many dust devils in Gale Crater. The formulation presented here provides a simple but robust scheme for relating the dust devils' statistical and physical properties, and it represents an important next step in improving our knowledge of these dynamic and ethereal phenomena.

\section*{Acknowledgments}
The authors acknowledge helpful conversations with Leming Qu and Paul Simmonds. This research was supported, in part, by the Idaho Space Grant Consortium.

\bibliography{biblio.bib}
\bibliographystyle{plainnat}

\newpage

\begin{figure}
\includegraphics[width=\textwidth]{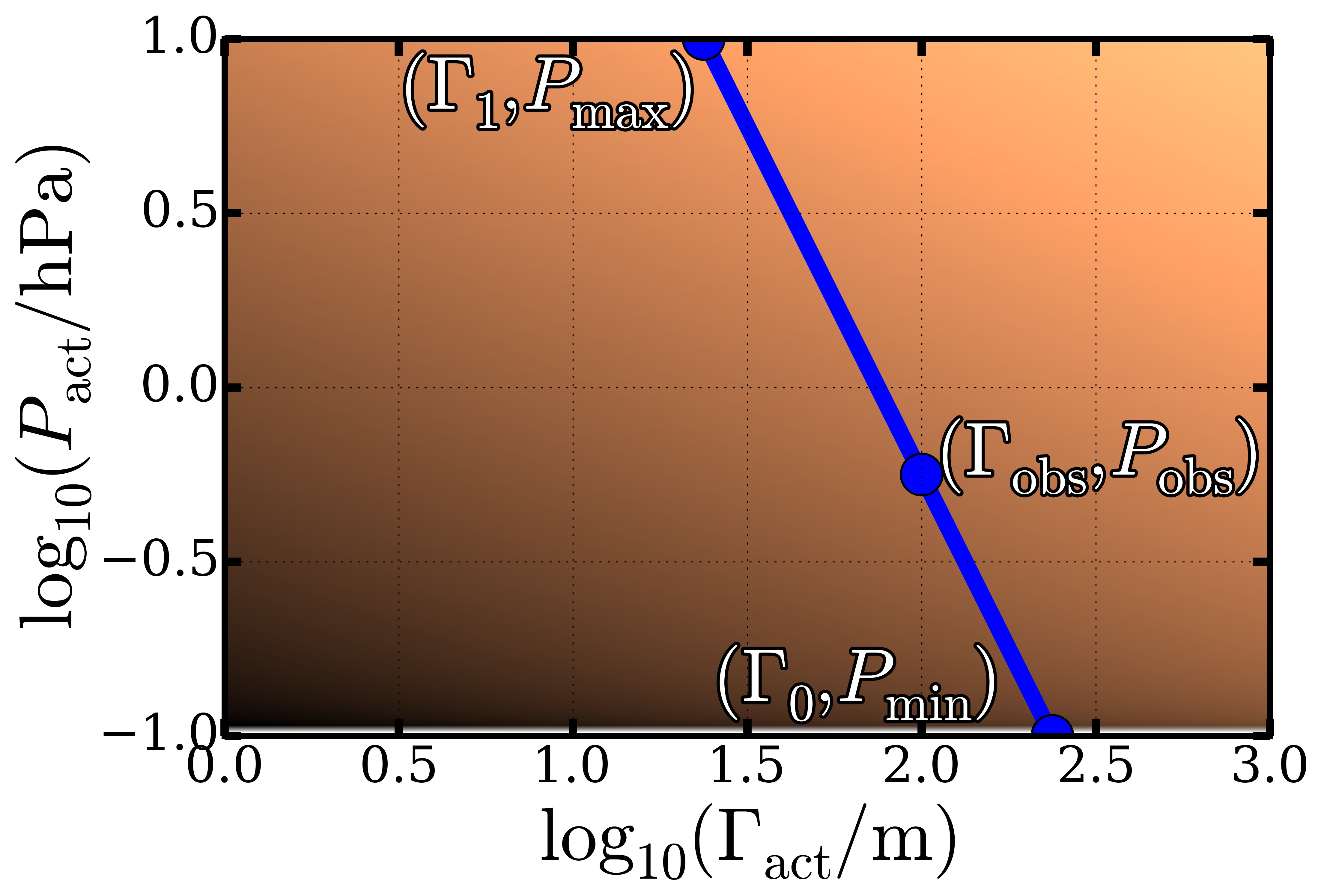}
\caption{Example of an integration track for Equation \ref{eqn:convert_from_actual_to_observed_density}. The shading illustrates an example $\rho(\Gamma_{\rm act}, P_{\rm act})$ distribution, discussed below -- brighter shades represent smaller densities. The blue track represents the locus of points passing through given $\Gamma_{\rm obs}$ and $P_{\rm obs}$ values. For example, $\rho(\Gamma_{\rm obs}, P_{\rm obs})$ involves integrating from $b = 0$, where $(\Gamma_{\rm act}, P_{\rm act}) = (\Gamma_{\rm obs}, P_{\rm obs})$ up to a maximum $b$ value, at $(\Gamma_{\rm act}, P_{\rm act}) = (\Gamma_1, P_{\rm max})$.}
\label{fig:integration_path}
\end{figure}

\appendix
\section*{Appendix}
\label{sec:appendix_a}

First, we provide the distribution of $\Gamma$ values described in Section \ref{sec:converting_durations_to_a_distribution_of_diameters}:

\begin{equation}
    \rho(\Gamma)= 
\begin{cases}
    k \alpha^{-1} \left[ \tau_{\rm min}^{-\alpha} - \tau_{\rm max}^{-\alpha} \right],& \text{if } \Gamma < \upsilon_{\rm max}\ \tau_{\rm min}\text{ \& } \Gamma < \upsilon_{\rm min}\ \tau_{\rm max}\\
    k \alpha^{-1} \left[ \tau_{\rm min}^{-\alpha} - \left( \Gamma/\upsilon_{\rm min}\right)^{-\alpha} \right],& \text{if } \Gamma < \upsilon_{\rm max}\ \tau_{\rm min}\text{ \& } \Gamma > \upsilon_{\rm min}\ \tau_{\rm max}\\
    k \left[ \left( \alpha^{-1} - \left( \frac{\upsilon_{\rm max} - \upsilon_{\rm min}}{\upsilon_{\rm max}} \right) \right) \left( \Gamma/\upsilon_{\rm max} \right)^{-\alpha} - \tau_{\rm max}^{-\alpha} \right],& \text{if } \Gamma > \upsilon_{\rm max}\ \tau_{\rm min}\text{ \& } \Gamma < \upsilon_{\rm min}\ \tau_{\rm max}\\
    k \left[ \alpha^{-1} \left( \Gamma/\upsilon_{\rm min} \right)^{-\alpha} - \left( \alpha^{-1} - \frac{\upsilon_{\rm max} - \upsilon_{\rm min}}{\upsilon_{\rm max}} \right) \left( \Gamma/\upsilon_{\rm max} \right)^{-\alpha} \right],& \text{if } \Gamma > \upsilon_{\rm max}\ \tau_{\rm min}\text{ \& } \Gamma > \upsilon_{\rm min}\ \tau_{\rm max}.
\end{cases}
\label{eqn:example_rho-gamma}
\end{equation}

Next, we discuss the details of Equation \ref{eqn:the_big_equation}. Figure \ref{fig:integration_path} illustrates the integration track involved in Equation \ref{eqn:convert_from_actual_to_observed_density}, and we define the end points of the integration track as $\Gamma_0 \equiv \Gamma_{\rm obs} \left( P_{\rm obs}/P_{\rm th} \right)^{1/2}$ and $\Gamma_1 \equiv \Gamma_{\rm obs} \left( P_{\rm obs}/P_{\rm max} \right)^{1/2}$. The equation involves an integral over $b$, which, given $\Gamma_{\rm obs}$ and $P_{\rm obs}$, represents a fixed curve in $\Gamma_{\rm act}-P_{\rm act}$. In other words, $\Gamma_{\rm obs}$ and $P_{\rm obs}$ define a locus of points for $\Gamma_{\rm act}$ and $P_{\rm act}$, and the integral over $b$ involves traveling along the locus from the point $(\Gamma_{\rm act}, P_{\rm act}) = (\Gamma_{\rm obs}, P_{\rm obs})$ up to $(\Gamma_1, P_{\rm max})$. In fact, any points $(\Gamma_{\rm obs}^\prime, P_{\rm obs}^\prime)$ satisfying Equation \ref{eqn:P_obs_Gamma_obs}, $P_{\rm obs}^\prime\ \Gamma_{\rm obs}^{\prime 2} = P_{\rm obs}\ \Gamma_{\rm obs}^2$, lie on this locus. Consequently, the only difference between $\rho(\Gamma_{\rm obs}^\prime, P_{\rm obs}^\prime)$ and $\rho(\Gamma_{\rm obs}, P_{\rm obs})$ is where on the track the integral starts -- the integrals for both end at the same point: 
\begin{eqnarray}
\label{eqn:difference_between_observed_density_points}
\rho(\Gamma_{\rm obs}, P_{\rm obs}) - \rho(\Gamma_0, P_{\rm th}) &=& \int_{(\Gamma_{\rm obs}, P_{\rm obs})}^{(\Gamma_1, P_{\rm max})} \cdots db^\prime - \int_{(\Gamma_0, P_{\rm th})}^{(\Gamma_1, P_{\rm max})} \cdots db^\prime \nonumber \\ 
& = & \int_{(\Gamma_0, P_{\rm th})}^{(\Gamma_{\rm obs}, P_{\rm obs})} \cdots db^\prime = \int_{b^\prime = 0}^{b} \cdots db^\prime,\end{eqnarray}
where we have suppressed the integrands for clarity. We can then differentiate both sides with respect to $b = \left( \Gamma_{\rm obs}/2\right) \left[ \left( P_{\rm obs} - P_{\rm th} \right)/P_{\rm th} \right]^{1/2}$, but, for the left-hand side, we will convert the $b$-derivative:
\begin{eqnarray}
\label{eqn:b_derivative_into_P_obs_derivative}
\frac{d}{db} & = & \left(\frac{dP_{\rm obs}}{db} \right) \frac{\partial }{\partial P_{\rm obs}} - \left(\frac{d\Gamma_{\rm obs}}{db} \right) \frac{\partial }{\partial \Gamma_{\rm obs}} \nonumber \\
& = & \left( \frac{2}{\Gamma_{\rm obs}} \right) \left( \frac{P_{\rm obs} - P_{\rm th}}{P_{\rm th}} \right)^{-1/2} \left( P_{\rm obs}\ \frac{\partial}{\partial P_{\rm obs}} - \left( \frac{\Gamma_{\rm obs}}{2} \right) \frac{\partial}{\partial \Gamma_{\rm obs}} \right).
\end{eqnarray}

Thus,
\begin{equation}
\label{eqn:convert_from_observed_to_actual_density}
\frac{d}{db} \bigg( \rho(\Gamma_{\rm obs}, P_{\rm obs}) - \rho(\Gamma_0, P_{\rm th}) \bigg) = \frac{d}{db} \left( \int_{b^\prime = 0}^{b} f\ \rho({\rm act}) \frac{2b^\prime\ db^\prime}{b_{\rm max}^2} \right),
\end{equation}
giving 
\begin{equation}
\label{eqn:appendix_the_big_equation}
\rho(\Gamma_{\rm act}, P_{\rm act}) =  k \Gamma_{\rm act}^{-5/3}\ \left( P_{\rm act} - P_{\rm th} \right)^{-1/2} \left( P_{\rm obs} \frac{\partial \rho(\Gamma_{\rm obs}, P_{\rm obs})}{\partial P_{\rm obs}} - \left( \frac{\Gamma_{\rm obs}}{2} \right) \frac{\partial \rho(\Gamma_{\rm obs}, P_{\rm obs})}{\partial \Gamma_{\rm obs}} \right)_{\rm obs \rightarrow act}.
\end{equation}

\label{lastpage}

\end{document}